\documentclass[pre,preprint,aps,]{revtex4}
\usepackage{graphicx,times}
\usepackage{comment}
\usepackage{amsmath,amssymb,amsthm}
\usepackage{bm}
\usepackage{verbatim}
\usepackage{float}
\usepackage[flushleft]{threeparttable}
\begin{document}

\title{The role of carbohydrates at the origin of homochirality in biosystems}

\author{S\o ren Toxvaerd }
\date{\today}

\begin{abstract}
Pasteur has demonstrated that the chiral components in a racemic mixture can separate in homochiral
crystals. But with a strong chiral discrimination the chiral  components in a concentrated
 mixture can also  phase separate into
 homochiral fluid domains, and the   isomerization kinetics can then
perform a symmetry breaking  into  one thermodynamical stable
 homochiral system. Glyceraldehyde
has a sufficient chiral discrimination to perform such a symmetry breaking.

 The requirement of a high concentration of the chiral reactant(s) 
in an aqueous solution  in order to perform and $\textit{maintain}$ homochirality;
the appearance of phosphorylation of almost all carbohydrates in the central machinery of life; the
basic ideas that the biochemistry and  the glycolysis and gluconeogenesis contain the trace of the biochemical evolution,
 all point in the direction of that
homochirality was obtained just after- or at a phosphorylation  of the very first products of the formose reaction,
at high concentrations of the reactants in  phosphate rich compartments  in submarine   hydrothermal vents.

A racemic solution of D,L-glyceraldehyde-3-phosphate could be the template for obtaining homochiral
D-glyceraldehyde-3-phosphate(aq) as well as L-amino acids.
\end{abstract}
 \maketitle
\section{Origin of life and the environment at Earth  4 billion year ago}

All carbohydrates and derivatives of carbohydrates in biosystems are
 D-configurations and
all amino acids and derivatives of are L-configurations. This notation refers to the
 stereo specific configuration of  ligands at the  carbon atom next to the
terminal  -CH$_2$OH for carbohydrates and to the configuration at the ($\alpha$-)
 carbon atom with a carboxylic acid and an  amine group  attached. The homochirality is crucial
for the coupled biochemical reactions in living systems.
The origin of this specific order could may be  have been evolved during a  purification of simple  biosystems, but
a general assumption is that the  homochirality  of carbohydrates and amino acids have been present  at the origin of life on Earth.

  The oldest undisputed evidence of life on Earth  is  3~Gyr (billion years) old
 \cite{Brasier}. It is fossilized bacteria, but older findings in rocks  dated to about 3.5~Gyr \cite{Altermann} and 3.8~Gyr \cite{Mojzsis}
 may also show sign of the presence of
bacteria.
 From a biochemical and biological point of view a bacteria is, however, a very complex and highly ordered system
and thus the purification of bio-molecules and the self-organisation into  complex biological control systems
 must necessarily have appeared significantly
earlier than the appearance of bacteria.
 The Earth is 4.56~Gyr old, and if life has originated on   Earth, it has taken place in  the
 time interval  from the creation of the Earth and to the appearance of the first geological sign of life. To determine possible
physicochemical models for the origin of homochirality and the origin of life it is thus necessary to know the
physicochemical environment at this early time of the Earth's history.

 Bacteria as well as all other biosystems  are
 from a physicochemical point of view soft condensed matter, and one more fact is worth noticing.
All the elementary  bio-assembly  of molecules are dehydrations
 (DNA,RNA,proteins,
starch,fats) with  release of one water molecule
per polymerization step, 
 with the consequence that H$_2$O must have been present at the
self-organization of the complex biosystems at the origin of life.
 A part of the   water in the oceans comes from meteors that contained ice  and
 hit the Earth  at the Late heavy bombardment 3.8--4~Gyr ago    \cite{Tera,Cohen}.
 But   the solid mantle of the Earth also contains
water \cite{Hirschmann} and there is evidence of oceans on Earth  4.4--4.3~Gyr long before the late heavy bombardment  \cite{Wilde,Mojzsis1}.
So a natural conclusion from these simple facts is that   life with homochirality
 of carbohydrates and amino acids  started as self-organization in a   fluid  containing water as one of its components. 

Another fact worth noticing is that all known  biosystems only exist at moderate temperatures,
  $T \leq 150^{\circ}$C, and there are
at least two reasons  that life hardly could have started just after the creation of the Earth.
 First the temperature just after the 
condensation of the materials, which has led to the creation of the planet(s)
 was  too high for living systems with  membranes. Secondly the Earth is believed  to have been involved
 in a collision 30--100 Myr (million years)
 just after its creation, 
a collision which created the Moon \cite{Canup1,Canup2}.  
This collision and the creation of the Moon  had  drastic consequences not only for the temperature, but also for the
condition at the surface of the Earth \cite{Zahnle}. The moon was at the start much closer to the Earth  
 and at that time a day was significant shorter. This is due to the tide waves which have pushed
 the Moon out to its present orbit and at the same time have damped the rotations of the Earth and the Moon.
 If  Earth had oceans $\approx4.3$--$4.4$\,Gyr ago
there must have been very strong tide waves and a turbulent atmosphere. The
extrapolation back to times just after  the Moon's creation is difficult \cite{Kagan,Bills}, but all  extrapolations
indicate  much stronger tide waves for $\approx$ 4--4.4~Gyr at  the origin of life.

 From a physicochemical point of
view biosystems are non-equilibrium and open systems,  and the refined
 self-organization must necessarily have taken place in
an environment with high concentrations of the basic constituents  and a smooth
  and continuous input  over very long period
 of time of the constituents for the biosystems,
which primarily are carbohydrates and amino acids.
 It is hard to imagine that this could have taken place in a
turbulent atmosphere- or in the oceans of the Earth.
 But the right conditions could have been present at- or below the
 hydrothermal vents  \cite{Shock,Russell}.

Before  analyzing   the role of carbohydrates at the origin of homochirality in biosystems we
need to  summarize the appearance of carbohydrides and amino acids at
the origin of life and their physicochemical
behavior (Section~2), as well as the basic  physicochemical-kinetics mechanism for
 obtaining  and $\textit{maintaining}$ homochirality (Section~3).

\section{Aqueous solutions of amino acids and  carbohydrates}
\subsection{Prebiotic appearance of  amino acids and carbohydrates on Earth}

\subsubsection{Amino acids}

There are at least  three sources to  prebiotic amino acids:  by lightning in the
atmosphere, injected into the Earth from the interstellar space, or already   present in the Earth's 
 upper mantle or  synthesized  from the released
materials at or below the vents.

The spectacular experiment(s) performed
by \cite{Miller} and \cite{MillerUrey} demonstrated that more than 20 different amino acids were produced by sparks ignition
in a prebiotic atmosphere of water, methane, ammonia and hydrogen. In the original experiments Miller focused on amino acids,
but later spark experiments demonstrated that also formaldehyde and other carbohydrates
 are synthesised  \cite{Miller1}. So the lightning
in this prebiotic atmosphere produced both
the  basic constituents in Earth's  biological systems.

The carbonaceous meteorites   contain amino acids and in a non-racemic proportion with a 
 very small excess of L-amino acids.
The composition of these meteorites is considered to be close
 to that of the solar nebula from which the solar system condensed. The two
most prominent examples are the  Murchison meteorite \cite{EngelNagy,BadaMiller}
 and the Murray meteorite \cite{CroninMoore}. From these meteorites compositional data
 and from an estimated impact rate of prebiotic meteorites, \cite{PierazzoChyba} suggested
 that meteorites
might have contributed significantly by  of the order up to $10^8$ kg per year
 to amino acids on Earth at the time where the life could had started.
But even with this injection of amino acids the concentration in the oceans must have been low.
Earth's oceans now  contain of the order $1.4 \times 10^{21}$ kg H$_2 $O.
The simple amino acid glycine is  observed in the interstellar space \cite{Kuan}, and  the interstellar space could have seeded  
the Earth with organic molecules at the origin of life \cite{Herbst}.

Amino acids can also be synthesized at conditions similar to the
 physicochemical conditions at or below the hydrothermal vents \cite{Amend,Honda,Huber}.
 If  the materials from which the solar system has condensed contained
 the same materials as the carbonaceous meteorites, there might have
 been a substantial amount of 
amino acids  in  the upper part of the mantle of the Earth  at the origin of life.

\subsubsection{Carbohydrates}

The formose reaction is the spontaneous  condensation of formaldehyde
 into carbohydrides,  which was discovered already in 1861 \cite{Butl},
and  the aldol-like mechanism is described by \cite{Bres}.
The two first steps  are
\begin{figure}[!h]
 \begin{center}
 \includegraphics[width=0.80\textwidth]{formose.epsi}
\end{center}
\end{figure}

\noindent
which gives a racemic mixture of D-glyceraldehyde and L-glyceraldehyde.
  Formaldehyde, CH$_{2}$O,  is  the simplest carbohydrate  and it can be  synthesized
  from carbon dioxide.
It is produced  in a reducing atmosphere containing CO$_{2}$ and methane \cite{Miller1}.
  The condensation into carbohydrides is catalyzed
 by not only amino acids \cite{Weber}; but also  by naturally
  aluminosilicates occurring at the
 alkaline hydrothermal vents \cite{Gabel,Martin}.

 Simple carbohydrates are also detected in the interstellar space \cite{Herbst}
which must have seeded the Earth, not only by amino acids but also with carbohydrates, and  simple carbohydrates 
(formaldehyde) might also have appeared in the upper part of  Earth's mantel at the origin of life. 

The formose reaction is well known and is believed to be the basic synthesis of bio-carbohydrates
  and related bio-organic molecules  \cite{Gabel,Wash,Kim,Jalbout,Lambert}.

\subsection{ Stability and  isomerization kinetics of  carbohydrates and amino acids}

Both carbohydrates and amino acids are soluble in water due to the hydrophilic -OH groups in  carbohydrates and
to the hydrophilic -NH$_2$ and -COOH in amino acids which performs hydrogen bindings with water molecules. Both
carbohydrates and amino acids are  stable, but configurational 
 unstable in the sense that they, by an isomerization kinetics, change configuration at an asymmetric carbon atom
(i.e. an carbon atom with  chemical bounds to four different atoms)

\subsubsection{Carbohydrates}

The  kinetics
 which  changes   the configuration at a chiral center in a carbohydrate is 
 a ``keto-enol''  kinetics. The simplest example is the  kinetics for D-glyceraldehyde (GLA) in 
equilibrium with L-glyceraldehyde
\begin{figure}[!h]
 \begin{center}
 \includegraphics[width=0.75\textwidth]{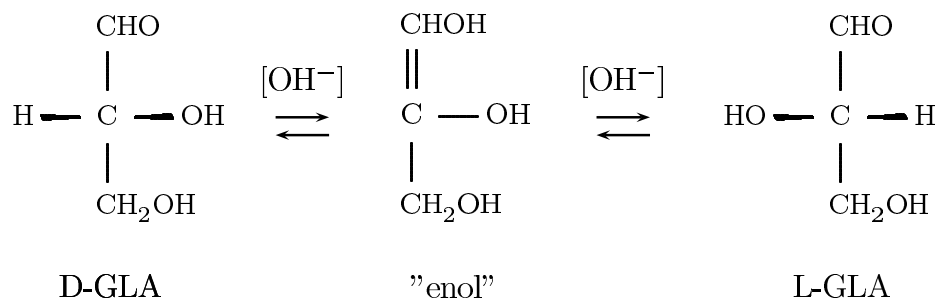}
\end{center}
\caption{Keto-enol isomerization kinetics for the triose glyceraldehyde}
\end{figure}

Glyceraldehyde has only one asymmetric carbon atom and the keto-enol kinetic converts the D-configuration to the mirror L-configuration.
The hexoses glucose and fructose have four and three asymmetric carbon atoms, respectively. 
The initiator of the configurational change by the keto-enol kinetics  is the oxo group at a carbon atom next to an asymmetric carbon atom,
 which at a hexose is either carbon atom No.~1 (e.g.\ glucose) or to carbon atom No.~2 (fructose).
The keto-enol intermediate is  a configuration with  two double bounded carbon atoms, each with an $-\textrm{OH}$ group, (see Figure~1). 
  It establishes a transformation between    different configurational
arrangements at   carbon atoms next to the oxo group, in the case of
  glucose  between   D-glucose, D-fructose and D-manose, but not 
between  the enantiomers D-glucose and L-glucose. The D-configuration of a carbohydrate
refers to the configuration of the asymmetric carbon atom next to the terminal group -CH$_2$OH,
in the case of glucose to carbon atom No. 5, and     $\textit{this is not involved in the keto-enol kinetics}$.
 To obtain a transformation between a D-carbohydrate and the  enantiomer  L-carbohydrate by the keto-enol  kinetics one must go to the triose, as demonstrated
in Figure~1.  L-hexoses  do not occur naturally in biosystems, but must be synthesized in the laboratory. (E.g.\ the acetate derivative
of L-glucose is found to stimulate insulin release, but it is produced synthetic \cite{Malaisse}).

Aqueous solutions of carbohydrates are reported to be unstable \cite{Kopetzki}. Reid and Orgel observed that diluted 
 solutions of carbohydrates under prebiotic conditions degrade within hours or days  \cite{ReidOrgel} and  
 ribose   decomposes  even at low temperature and neutral pH (44 years at 0$<$
  $T \leq 150 $  $^{\circ}$C)  \cite{Larralde}. None of the experiments contain, however, any information
about the quality of the water, used for the solutions. Even highly purified water contains 
organic materials \footnote{see www.elgalabwater.com/water-quality-en.html}
and it requires special laboratories and technique to avoid bacteria in a solution.
Bacteria, or their enzymes will  degrade carbohydrates at low concentrations, but not at high concentrations. Carbohydrates at high concentrations are
known to be very stable and they are used for conservation of food.
Honey, which can be  characterized as a  fructose rich-solution with $\approx 17\%$ water is very stable,
and fossil honey  was found in in southern Georgia in a tomb from the bronze age, 27th--25th centuries B.C.\ \cite{honey}.

\subsubsection{Amino acids}

Amino acids undergo also an isomerization kinetics \cite{Bada}, but  there are   crucial differences between the
keto-enol kinetics for glyceraldehyde, pentoses and hexoses  on one hand
 and the isomerization kinetics for  bio-amino acids on the other hand. The  pentoses and hexoses maintain their
D-configuration at the keto-enol kinetics and only the carbohydrates with one asymmetric carbon, glyceraldehyde,
 changes between the D- and L-configuration. But
 all  the chiral bio-amino acids change the L-configuration to the D-configuration by their isomerization kinetics.
This is because  all bio-amino acids are $\alpha$-amino acids, i. e. the D- and L- configurations refer to the  asymmetric carbon atom with
an amino and a carboxylic group attached.

 The isomerization kinetics will drive a 
 diluted solution of a homochiral L-amino acid toward
a racemic solution \cite{Bada}. Another difference between carbohydrates and amino acids is
 the values of the reaction constants for
 the isomerization kinetics. Amino acids change configurations within thousands of years \cite{Bada}
(but might racemize much faster in the presence of a dipeptide \cite{Boehm}).
Glyceraldehyde changes its configuration within hours, but with competing reactions \cite{Fedoro}.

 Although the amino acids racemize they are, however, very (constitutional) stable in solid
 phases as well as in solutions, and even
at relative high temperatures (423 K) \cite{Conway,Snider}. So if
 the composition of the carbonaceous meteorites equals the composition of the materials
that created the Earth it is actual possible that the mantle of
the Earth has contained amino acids at the origin homochirality and  life.

\section{ Spontaneous symmetry breaking in concentrated racemic solutions }

Pasteur has demonstrated that the chiral components in a racemic mixture can separate into
homochiral crystals of each enantiomers.
A racemic system can, however,  perform a spontaneous symmetry breaking
  to $\textit{one}$ homochiral system. There are $\textit{two}$ necessary conditions for that
 this process  takes place. The reaction must be exergonic, i,e.
$\Delta_r G<0$ and  the system must have an   isomerization kinetics  between
the two chiral configurations. Some amino acids may fulfill these two conditions, but in the case of
carbohydrates only the 
triose glyceraldehyde and its derivatives, e.g.\ glyceraldehyde-3-phosphate fulfills it. This is due to
the fact that whereas the isomerization kinetics for $\alpha$-amino acids converts D- to L- and visa verse,
the keto-enol kinetics for higher-order carbohydrates as mentioned does not change the stereo configuration at the carbon atom next
to the terminal group.

\subsection{Exergonic reaction in a racemic mixture of chiral molecules with  strong chiral discrimination}

The first and crucial condition for obtaining not only a spontaneous symmetry breaking,
 but also for maintaining homochirality over long times
is that there is a (exergonic) gain of  Gibbs free energy by a  separation into homochiral sub domains.
For the reaction 
\begin{equation}
 D,L \rightarrow D+L,
\end{equation}
where D,L is an uniform mixture of molecules with D- and L configurations 
 which separates into  $\textit{sub domains}$ of the D and L forms, the
 the gain in Gibbs free energy, 
\begin{equation}
\Delta_r G = \Delta_r H-T\Delta_r S
\end{equation}
by this phase separation 
is primarily determined by the gain in enthalpy, $ \Delta_r H < 0$ by the separation. This is because the
change in free energy from the entropy term by the separation, $-T\Delta_r S>0$,
is positive and of the order $RT\ln 2 \approx2$\,kJ/mole for a racemic mixture of equal amount of
molecules with
D- and L configurations 
(obtained by approximating the D,L mixture with an ideal mixture with mixing entropy equal
to $R[x\ln x+(1-x)\ln(1-x)]$ and with mole fraction $x=1/2$ for the racemic mixture).

 The  gain in enthalpy is given by    the strength of the
chiral discrimination (cd).
The existence of chiral discrimination was discovered by Louis Pasteur in 1848 \cite{Pasteur}. 
It manifests itself  in the crystallization from a  racemic
fluid mixture of  enantiomers into
crystals of homochiral molecules, as demonstrated by Pasteur.
As is well known, it is due to the fact that for some molecules there is
an enthalpy gain, $-\Delta_r H>0$, by packing homochiral
 molecules together instead of in pairs of enantiomers. The enthalpy  gain must be bigger
 than the temperature times the  mixing entropy, which is of the order  2\,kJ/mole
  at room temperature.
  For pure systems, $ \Delta_r H$  in Eq.~(2)  for the change in enthalpy by the
phase separation due to the chiral discrimination   can
be obtained from standard values as
    \begin{equation}
   \Delta_r H=   \Delta_{\textrm{cd}}H^{^{\ominus}}= \Delta_{\textrm{f}}H^{^{\ominus}}({\rm enantiomer})
         -\Delta_{\textrm{f}}H^{^{\ominus}}({\rm racemic})
	     \end{equation}
From a physicochemical point of view 	     
biosystems are, however, in an aqueous solution. Because the polymerization of amino acids, as well as of carbohydrides and Nucleic acids,
are dehydrations with a release of one water molecule per
covalent polymerization bond, the state where the synthesis of prebiotic systems appear
is, what in  physical chemistry is referred to by (aq) 
for the aqueous solution.
On the other hand, the concentration of the chiral molecules, $[D]$ and $[L]$, at  prebiotic self-assembling must necessarily
have been  high, with only a small concentration of water molecules, for two  simple
reasons. First
\begin{equation}
\Delta_{\textrm{cd}}H \approx 0; [D], [L] \ll 1
\end{equation}
 in diluted solution because there is no aggregation of homochiral clusters,
so the reaction is endergonic ($\Delta_r G \approx -T\Delta_r S>0$), and secondly 
  the isomerization kinetics (the second general  condition) in a diluted aqueous solution drives the
system toward a racemic composition, as is observed in diluted aqueous solutions of L-amino acids \cite{Bada}. So even
if one, in one way or the other, starts with a homochiral system, it will  tend to a racemic state
if the concentration is low.

\begin{table}[htb]\caption{
Strength of chiral discrimination for different molecules.
}
\vspace{12pt}\centerline{\begin{tabular}{lc}
Molecule & $\Delta_{\textrm{cd}}H^{^{\ominus}}$/kJ\,mol$^{-1}$ \\
\hline
    Glyceraldehyde(l) &  -- 11.7 \\
    Phenylalanine(c) &   -- 6.7 \\
    Isoleucine(c) &   -- 5.4 \\
   Lysine(c)   & $\approx$ 0. \\
   Valine(c)   & $\approx$ 0. \\
   Leucine(c)  &     3.1 \\
   Serine(c)   &  6.3 \\
 Proline(c)    &   16.6 \\
 Threonine(c)  &    16.9 \\
 Alanine(c)    &    18.9 
\label{Table}\end{tabular}}
\begin{tablenotes}
\small
\item 
Data taken from NIST Standard Reference Database Number 69 and references therein.
\end{tablenotes}
\end{table}

The difference between Pasteur's observation of the spontaneous separation of enantiomers from a racemic mixture
and a corresponding spontaneous separation for a racemic mixture of carbohydrates or amino acids is
that Pasteur's chiral constituents: D- and L-ammonium sodium tartrate are moderately soluble in water, but
carbohydrates and amino acids are very soluble. But even if it is possible to obtain a separation of e.g.\ amino acids
 by a crystallization, biosystems are soft condensed
matter and we need to find a mechanism which not only separates
the two constituents in the racemic composition, but also stabilizes homochirality over (very) long times in  a
 water environment.
 Therefore a  measure of the strength of the chiral discrimination should
be obtained for a state with high concentration of the chiral molecules  and where
\begin{equation}
 \Delta_r H \approx \Delta_{\textrm{cd}}H^{^{\ominus}}(l),
\end{equation}
and this  should at least be bigger (i.e. $\Delta_r H< -2$\,kJ/mole) than the
 mixing entropy to ensure an exergonic process.

 In general,   physicochemical data for the strength of chiral discrimination,
  $ \Delta_{\textrm{cd}}H^{^{\ominus}}(l)$ in the fluid state do not exist.
  Only for the simple triose glyceraldehyde (Gla)
   where  $ \Delta_{\textrm{cd}}H^{^{\ominus}}(l)=-11.7$\,kJ/mole \cite{Gla}, which is much more than
     what is needed to overcome the mixing entropy. For some amino acids
      there exist data for  $ \Delta_{\textrm{cd}}H^{^{\ominus}}(c)$ for the enthalpy difference between
       an enantiomer crystal and a racemic crystal. This strength is, however, weakened at melting
        due to  less effective packing in the fluid state. A collection of available data is given in Table~1.
As can be seen from this table a few central amino acids (Phenylalanine, Isoleucine)  might have the potential to maintain
a homochiral  aqueous state (aq). The simple triose, glyceraldehyde  has, however, an
exceptionally  and sufficiently strong chiral discrimination.

\subsection{The isomerization kinetics}
The second general condition for obtaining a spontaneous symmetry breaking
in a racemic mixture  is the presence of an isomerization kinetics between
the two enantiomeric configurations.
The isomerization kinetics of amino acids and carbohydrides are complex
\cite{Bada,Fedoro,Nagor}. We shall
simplify the reaction mechanism  by bimolecular reactions 
\begin{equation}
\begin{array}{lllll}
 & E_{\mathrm{DD}}& & E_{\mathrm{DL}} \\
  \textrm{D + D}& \rightleftharpoons
   &\textrm{D + L}& \rightleftharpoons& \textrm{L + L}\\
    & E_{\mathrm{DL}}& & E_{\mathrm{LL}} \\
    \end{array}
    \end{equation}
    between the two chiral species.
  The activation energy, $E_{\mathrm{DL}}$, for   DL-collisions in an exergonic reaction,
         which may  convert  a  D-molecule
	  into  a L-molecule or \textit{vice versa},
	  is, in a condensed racemic fluid,  less than the corresponding activation energy,  $E_{\mathrm{DD}}
	     = E_{\mathrm{LL}}$, thus allowing a conversion  of one of the molecules in  the collisions.
The     inequality  
\begin{equation}
E_{\mathrm{DL}}<E_{\mathrm{DD}}= E_{\mathrm{LL}}
\end{equation}
accounts for the chiral discrimination with
 a lower potential energy   in homochiral domains of one of the enantiomers corresponding to a sufficient
 strength of the chiral discrimination \cite{Atkins}.
 More specifically
\begin{equation}
E_{\mathrm{DL}} +  \mid\Delta_r H(x)\mid =  E_{\mathrm{DD}}= E_{\mathrm{LL}},
\end{equation}
where the gain in enthalpy, $\mid\Delta_rH(x)\mid$, by a conversion of the stereo configuration  is a function of the 
$\textit local$ composition, $x({\bf r})$ of the configurations  at the position  ${\bf r}$,
 and with a maximum (negative)  value of $\mid\Delta_rH(x)\mid$ 
 for a homochiral local composition ($x({\bf r})=0$ or $x({\bf r})=1$).

 Primarily   $\mid \Delta_rH(x) \mid$
is proportional with the excess number of homochiral neighbors by a change of a configuration. Consider a simple example:
Let a molecule, e.g.\ in a L configuration be activated to  a transition state (intramolecular) configuration.
 It will with a Boltzmann probability
end in the configuration with lowest energy. A simple molecule has twelve nearest neighbours. Let e.g.\ five of them be
in a D configuration, four of the neighbours in a L-configuration and two be  indifferent
water molecules. It corresponds to that the
local environment before the molecule was activated was racemic with  an equal amount of D and L configurations. But the activated
 molecule will most likely  turn into a D configuration by which the 
system tends to a lower energy, but now with an excess of D configuration. Thus
a strong chiral discrimination  together with an  isomerization kinetics will ensure a  separation
of a racemic mixture on a molecular level and
 tend to order the molecules in homochiral clusters and sub domains.

A racemic D,L-system tends to separate in homochiral domains of enantiomers if the enthalpy gain,
due to the chiral discrimination, is bigger than the entropy of mixing, just as in the case of Pasteur's experiment,
but now in a fluid state. 
Without  an isomerization kinetics such a racemic mixture will  separate into homochiral sub domains  through molecular diffusion.  
 An  isomerization kinetics enhances the separation;
 but what is much more significant, the kinetics also ensures a break of symmetry  as demonstrated by the simple example, and  results
			 in the dominance of one of the species   at late times \cite{tox1}.

The break of symmetry is obtained by, what could be expressed as
{\textit{self-stabilizing chance}} \cite{toxb}. The deviation from a racemic mixture is self-stabilizing because
 a homochiral clusters- or sub domains catalyse their own growth by the isomerization kinetics
 which mainly takes place in the interface, whereas  the chiral discrimination
 inside the homochiral domains slows down the kinetics and the conversions of the configurations, which always are unfavorable.  
 Still one needs to explain the observed break of symmetry since the kinetics 
seems only to enhance the separation but $\textit{it does not  favour one of the chiral}$
$\textit{ species}$.
The break of symmetry  on a macroscopic scale will appear when one of
the homochiral domains encapsulates the other domain. In a \textit{confined geometry} of e.g.\ a
fluid in a bottle or a prebiotic racemic mixture in a solid chamber \cite{Russell2,Russell1}, one of the homochiral
domains will dominate when, by a chance, it  percolates in the confined volume and
encapsulates the  other domain.

\begin{figure}[!h]
\begin{center}
\includegraphics[width=60mm,angle=-90]{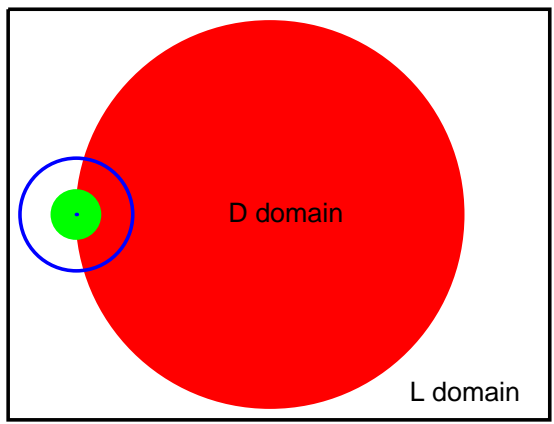}
\end{center}
\caption{Illustrative example of the  topological mechanism for  symmetry breaking and the dominance of one of the
chiral species. The (rectangular) volume contains the two separated homochiral phases of the enantiomers. The L-rich domain
is white and it percolates the volume, whereas the red D-rich domain in the droplet is encapsulated by the L-rich domain. An
activated particle (green) in the interface between the two domains has a range of attraction (blue) to the nearest neighbours which
together with the activated particle determine the strength of the chiral discrimination.
 Since the white area inside the blue circle is greater than the corresponding red area,
there are an excess of  L neighbours and the
activated particle will end in the L-configuration with a Boltzmann probability.
 The white area will increase and the area of the red droplet decrease and the droplet will finally disappear. }
\end{figure}

 A simple example (Figure~2) illustrates this mechanism.
Consider a system consisting of two homochiral (macroscopic) phases and let e.g.\ the D form
be the dominating (red area in Figure~2), but  as a big droplet in the middle of a box surrounded by
molecules with  L configurations (white area). The  D molecules  in the droplet has a convex surface ( and 
 the L molecules  around the droplet have correspondingly a concave surface).
Let the green circle mark an activated molecule $A^*$ which will tend to either a D configuration or
a L configuration with a Boltzmann probability depending on the local strength of $\Delta_rH(x(\textbf{r}))$.
  Due to the strong chiral discrimination
$\Delta_rH(x(\textbf{r}))$ is only negative in the interface and the
isomerization kinetics is mainly active there.
Even if there are more  D molecules in
the droplet  than L molecules around the droplet, most neighbours to an activated molecule in
the  interface are L due to the concave L-rich neighbourhood, and the activated molecules will most likely
end in a L configuration by which the (big) droplet decreases. 
  It is a matter of {\textit{ chance}} which
of the two enantiomers  that dominates and encapsulates the other domain.
The  domain-catalytic behaviour
is {\textit{self-stabilizing}}, and the dominance of a percolating domain is
self-stabilizing as well, because  its concave interface will ensure the dominance. The topological determined condition for 
symmetry breaking and the dominance of one of the species is illustrated in Figure~2 \cite[for further details see][]{tox1}.

Figure~3 shows the time evolution of chiral dominance in a fluid
of particles with chiral discrimination, obtained by a realistic (molecular dynamics)
computer simulation \cite{tox1}. The kinetics are implemented at time $t=0$ and
the strength of the chiral discrimination is given
by the difference in the two activation energies, $E_{{\mathrm{DL}}}$ and $E_{{\mathrm{DD}}}=E_{{\mathrm{LL}}}$.
The dominance of a species is measured by the enantiomeric excess given by the difference in mole fraction, $x_{{\mathrm{D}}}-x_{{\mathrm{L}}}$.
For  $E_{{\mathrm{DD}}}-E_{{\mathrm{DL}}}=RT$ we do not obtain a break of symmetry even after very long times,
and inspection of the (local) composition shows that the mixture remains in the racemic state.
This is also to be expected since the mixing entropy is  $RT\ln 2$. But for a stronger chiral discrimination
one observes a break of symmetry and a dominance of one of the enantiomers at late times. 
The figure  only shows four cases; but many  independent simulations gave an
equal dominance of the D- and L- systems, within the statistical uncertainties, as one
shall demand for a racemic system with no preference for one of the enantiomeric configurations.
The dominance appears
already for a strength of 2$RT$ (b in the figure), but this strength is not sufficient to kill the isomerization
kinetics within the dominating domain, and the system ends in a chiral fluid,
but with a small  homogeneous concentration of the loosing species. (Only for an infinite
strong chiral dominance of  $ \Delta_rH(cd)
 = \infty$ (d in the figure)
does this system end in a pure homochiral fluid.) Two other fluid systems have been investigated
\cite{tox2,tox3} in order to determine the sensitivity of the break of symmetry of the actual
fluid systems,  the conclusion being, that the isomerization kinetics which destabilizes
homochirality in diluted solutions  ensures homochirality at high concentrations, provided that the
chiral discrimination is sufficiently strong.

\begin{figure}[!h]
\begin{center}
 \includegraphics[width=0.60\textwidth]{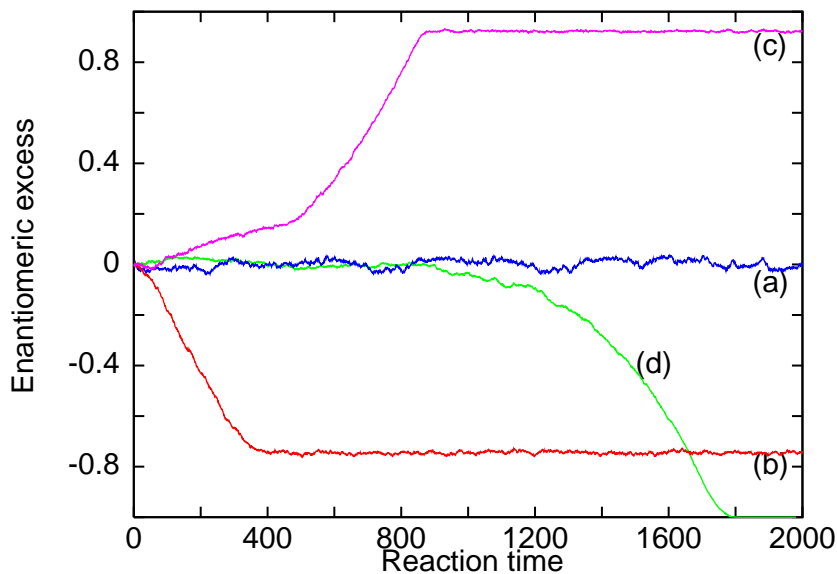}
\end{center}
\caption{ The enantiomeric excess, i.e the difference in mole fraction, $x_{{\textrm{D}}}-x_{{\textrm{L}}}$, as a function of
time in a molecular dynamics system with isomerization kinetics and for different
strength of chiral discrimination, accordingly to Eq.~(8): (a)  $\mid \Delta_{cd}H^{^{\ominus}} \mid =RT$,
(b) $2RT$, (c)  $3RT$ and (d) $ \infty$.}
\end{figure}

The description of the spontaneous symmetry breaking is exact. It makes only use of the law of thermodynamics and
that  the  kinetics is reversible. But it is formulated for
an isomerization kinetics between the two enantiomeric configurations in  a concentrated
aqueous solution.
 There are other models for  symmetry breaking
 and spontaneous homochirality in racemic systems.
Some of these models differs 
 by assuming that the low  Gibbs free energy state is not a
 concentrated aqueous solution of the basic constituent, but
 e.g.\  the crystalline
phase or adsorbed constituents at mineral surfaces \cite{Hazen}.
 The general condition cited above is of cause also valid for these models,
but the reason for focusing on concentrated aqueous solutions is  that
it is not enough to obtain a homochirality  by crystallization or through  adsorption at a mineral surface. A bacterium
 is, from a physicochemical
point of view, soft condensed matter containing    chiral molecules including  both the 
L-amino acids and simple D-carbohydrates in an aqueous solution \cite{Ellis,Spitzer}, and the nucleus for life must
from a physicochemical
point of view, have been an open driven system with a steady state input  of the
 chiral units for the 
self-assembly.
So  the homochirality must be maintained in this fluid  over a very long period of time  at the 
 emergence of a living system.

In other models  for the origin of homochirality in
biosystems the chiral purification is obtained through the consecutive
kinetic equations for the polymerization of the complex organic molecules. In the Frank model \cite{Frank,Brandenburg} the composition of
 the simple chiral units in the biopolymers 
is  assumed to be racemic at the origin of the synthesis, and the overall homochirality is obtained
by the kinetics
during a period of time at the origin of life.  The homochiral domains are here
 the polymers and the chiral
discrimination is given by the Gibbs free energy difference between the formation of a polymer binding
 between two homochiral centre
 versus the corresponding energy of a polymer binding between units with L- and D- configurations. This asymmetry is  expressed by the
rate constants for cross-inhibition.
 This model is not at all in conflict with the present
description. In both cases  it is the kinetics in the open driven system which ensures homochirality, either by the
isomerization kinetics between monomers or by the kinetics of polymerization, and in both cases governed by
the gain of Gibbs free energy  by  homochiral  (inter- or intramolecular) contacts.
In fact both  symmetry breaking kinetics might have been present at the origin of homochirality. But whereas the Frank model
 is an appealing model for purification
of  a racemic mixture of amino acids, the  model is less  appealing in the case of homochirality in  polymers of carbohydrates such
as  the complex stereo-specific macromolecules DNA and RNA with a homochiral backbone of D-ribose. Here a model which ensures
homochirality of the polymer units before the synthesis of the macromolecule is more appealing.  

\section{Homochirality of carbohydrates and amino acids }

The carbohydrates, amino acids and their derivatives in biosystems are all homochiral.
 Thus one needs to explain  how a symmetry breaking appears
for both constituents.  No matter where the ordering has taken place it is  likely that
both constituents  have been present. But it is in fact very simple to explain this apparent puzzle.
The mechanism works primarily through the gain of Gibbs free energy produced by a homochiral clustering together with
an  isomerization kinetics,  and  works equally well for mixtures of carbohydrates and amino acids provided there
is a sufficient gain in energy by clustering D-carbohydrate molecules together with L-amino acid molecules. Nowhere
in the thermodynamic/kinetic description in Section 3 for  symmetry breaking in
a concentrated aqueous solution is it assumed that all the molecules should be of the same type, only that there
is a gain in energy by ordering and an active isomerization kinetics in the mixture. It is
only a matter of whether there is a sufficient gain of Gibbs free energy
 by clustering amino acids together with a carbohydrate and this
is actually the case for some simple carbohydrates and some amino acids \cite{Kock,Takats,Sandra1,Sandra2,BreslowCheng}.

In \cite{Kock,Takats} the authors observed a stereo selective coupling between L-serine octamers and D-glyceraldehyde and D-glucose.
Serine furthermore performs   coupling with phosphoric acid and some transition-metal ions.
Serine has a unique ability to perform "magic-number" ionic clusters composed of eight acid units and with a stereo selective 
chiral discrimination \cite{Serine}.

L-isovaline  acts as a asymmetric catalyse of the  threose and erythose from condensation of glycolaldehyde \cite{Sandra1}.
Glycolaldehyde and racemic glyceraldehyde condensate to pentoses in non racemic proportions  in the
presence of (some) dipeptides of L,L- and D,D-amino acids \cite{Sandra2}.

 The syntheses of glyceraldehyde from
formaldehyde and glycolaldehyde in the present of homochiral amino acids gives a non racemic yield \cite{BreslowCheng}.
In \cite{BreslowCheng} some L-amino acids were observed to give an
 excess amount of D-glyceraldehyde, e.g.\ L-phenylalanine, whereas e.g.\
L-proline gives a significant excess of L-glyceraldehyde.

These  asymmetries, obtained in the synthesis of carbohydrates in the presence of excess L-amino acids, must be a result of the
chiral discrimination (energy binding)  between the carbohydrates and  the amino acids. According to the model
described above such  gain of Gibbs free energy 
 together with the  isomerization kinetics of glyceraldehyde is sufficient to obtain
a homochiral solution of D-glyceraldehyde. But the available experimental data show that
 only some of the L-amino acids favor an ordering
of D,L-glyceraldehyde to D-glyceraldehyde. For some amino acids their L-configurations
leads to an excess of L-glyceraldehyde \cite{Sandra2,BreslowCheng}, e.g.\ L-proline. These
catalysis works both ways so an excess of L-glyceraldehyde  enhances the isomerization growth of
L-proline, and correspondingly  D-glyceraldehyde must enhance the isomerization growth of
D-proline. According to Table~1 proline has a chiral discrimination which favors a
racemic proportion even in a concentrated solution.  Thus if the origin of homochirality of carbohydrates appeared via a catalyze by
an L-amino acid, one needs to clarify how the homochirality of some  L-amino acids, e.g.\ L-proline, has been obtained together with D-carbohydrates.

The chiral discrimination between  amino acids and glyceraldehyde do not favour one of the constituents.
But there is, however a significant difference between the isomerization kinetics of a carbohydrate and an amino acid.
Whereas a diluted solution of an amino acid racemizes within thousand of years  \cite{Bada} \cite[see however][]{Boehm}, the corresponding racemization of
e.g.\ a diluted solution of D-glyceraldehyde is significant faster  but with competing reactions \cite{Fedoro}. On one hand this 
 may be taken to imply  that it is only possible within days or years to observe an excess of chirality of  carbohydrates,
 but not of amino acids in a mixture
of the two constituents, and one the other hand,  that homochirality is first obtained
 for the  carbohydrate no matter what the excess of chirality of the two constituents at
the start of the purification. 

\section{The role of phosphate at the origin of life}

So far one important  fact has been ignored. The role of phosphate at the origin of homochirality and life.
Although all derivatives  of carbohydrates have D-configurations they appear almost always as phosphate derivatives.
The list is long: The central energy cycles, the glycolysis and gluconeogenesis, ATP, ADP, RNA, DNA and the most common anchor molecule in the
membrane molecules. 
Phosphates did not, however, appear in high concentrations in the early oceans \cite{Keefe,Hagan}, but they are widely distributed in
 many minerals.
 Phosphate rocks containing apatite ( Ca$_{10}$(PO$_{4}$)$_{6}$(OH,F,Cl)$_{2}$), are
widespread on Earth and there are rare natural
examples of condensed phosphates (e.g.\ canaphite,
Ca$_{2}$Na$_{2}$PO$_{7}$.4H$_{2}$O \cite{Rouse}; Kanonerovite MnNa$_{3}$P$_{3}$O$_{10}$.12H$_{2}$O \cite{Popova}).

 A simple example illustrates the role of  phosphate at the origin of life.
There exist two different forms of simple living organism, Archaea and Bacteria. The central metabolisms (glycolysis and gluconeogenesis)
are common but 
 they differ on other biosynthetic pathways \cite{Graham,Genschel}, and by having different membrane molecules.
 Archaea membrane-molecules  consist of (ether) derivatives of sn-glycerol-1-phosphate,
whereas the molecules in the membranes of Bacteria and Eucarya 
 consist of (ester) derivatives of sn-glycerol-3-phosphate.
The membrane molecules are now synthesized by enzymes which, however,
 have no common  sequence, and
 an analysis of these  enzymes in different
  Achaea and Bacteria indicates that there is no common enzyme-ancestor
 \cite{Koga}. 
In a living organism the  membrane building blocks  are  synthesized from 
e.g.\ glycerol with help of these stereospecifique enzymes \cite{Pere};
but at the origin of life we must search for a reaction pathway
of spontaneous self-assembling directly from the ingredients.
There are two facts which together are  notable.

\begin{figure}[!h]
 \begin{center}
 \includegraphics[width=0.80\textwidth]{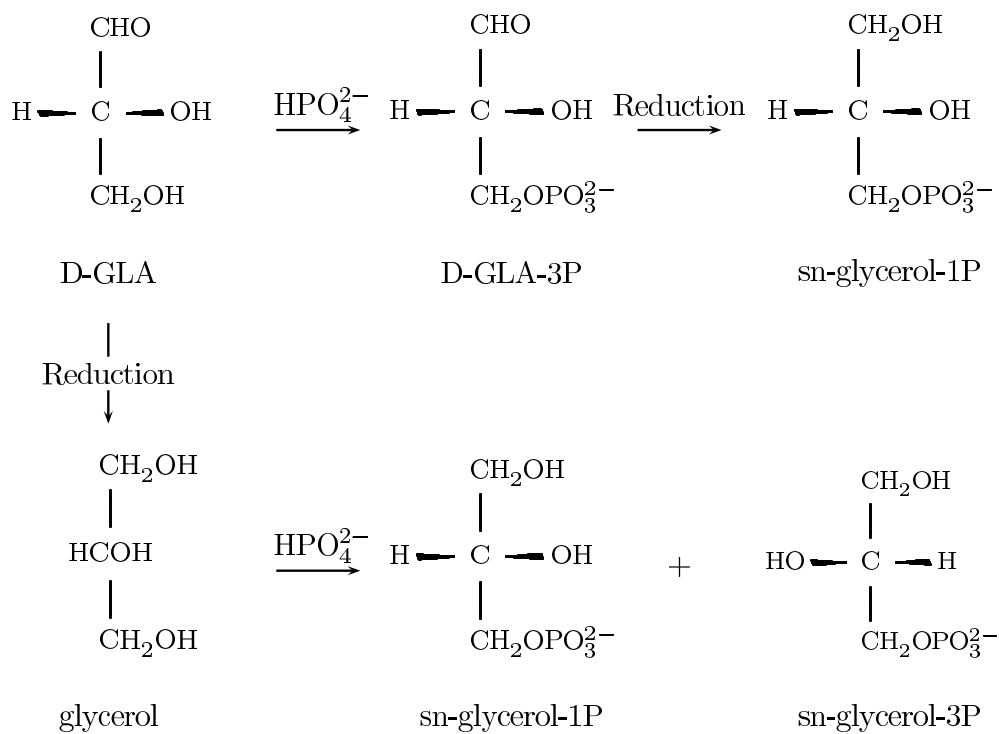}
\label{invsimpl}
\caption{ Phosphorylation and reduction of glyceraldehyde. If phosphorylation precedes reduction one obtains
the configuration found in Archaea;  if the reduction  precedes  phosphorylation 
one obtains both the configurations in Archaea and Bacteria. (Notice that due to $IUPAC$ nomenclature rules the reduced product of
D-glyceraldehyde-3-phosphate is sn-glycerol-1-phosphate.)}
\end{center}
\end{figure}

All the elementary  membrane molecules in a cell  are homochiral,
 and all  these  molecules  (in Archaea and Bacteria membrane
of three carbon atoms)  can be synthesized from glyceraldehyde or glycerol.
These two facts are  notable because one can obtain simple vesicles with membranes
 from amphiphiles with only one or two central
carbon atoms   and  with only  one  hydrophobic chain in 
 the membrane molecule \cite{Ouri}.
 But nature did it with three
carbon atoms in the central  chiral unit
 and with two hydrophobic chains linked to the amphiphilic head group;
all which can be synthesized
from glyceraldehyde  (see Figure 4).
 If the spontaneous self-assembly  of cell-membranes  
   had originated  from a pool of simple bio-molecules, why would the spontaneous
 synthesis not have chosen the simplest track and synthesized
 simpler  membrane molecules than diglycerides,  phosphoro-lipids and phosphoro-ethers?
The homochirality of the biomembranes are crucial for their functions and
  the first chiral component in
 the formose reaction is glyceraldehyde from which 
D-glyceraldehyde-3-phosphate and all the different membrane molecules can be synthesized.

\section{ The role of carbohydrates at the origin of homochirality in biosystems}

The general opinion and the literature about the origin of homochirality favors amino acids as the initiator and catalyzer
of homochirality \cite{Higgs}. But inspection of the experiments and the present analysis reveals that it is more likely
 D,L-glyceraldehyde or rather D,L-glyceraldehyde-3-phosphate.
The catalysis of  symmetry breaking by the chiral discrimination of carbohydrates and amino acids
 is neutral in the sense that it does
not select one of the reactants, e.g.\ an amino acid, as the catalyzer and the other as the substrate. A
 differences in the reaction times of the isomerization kinetics in favor of carbohydrates implies, however,
 that a carbohydrate reaches the homochiral state
 faster, and thus will act as the catalyzing template for homochiral purification of the amino acids.
 But if it was a carbohydrate that first reached the homochiral state, there seems to be only one possibility, glyceraldehyde
or more likely the derivative  glyceraldehyde-3-phosphate. This is so  because
the  isomerization kinetics of the higher order carbohydrates  do not change the
configuration at the carbon atom next to the terminal group, and because the spontaneous
reaction of D,L-glyceraldehyde into pentoses and hexoses results in a spectrum of carbohydrates
  without an enantiomeric excess \cite{Weberformose}.
These fact  favors a phosphorylation before or at a symmetry breaking and with the homochirality
of D-sugars and L-amino acids obtained
in a concentrated aqueous solution of D,L-glyceraldehyde-3-phosphate.

\begin{figure}
 \begin{center}
\includegraphics[width=100mm,angle=0]{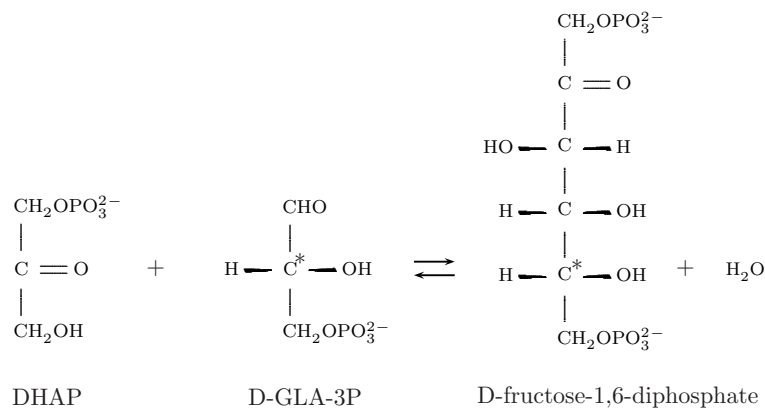}
\label{invsimpl}
\caption{ The aldol condensation of dihydroxy acetone phosphate (DHAP) with D-glyceraldehyde-3-phosphate
(D-GLA-3P) to D-fructose-1,6-diphosphate. The asymmetric carbon atom  from D-GLA-3P which,  by the condensation gives the D configuration in
  D-fructose-1,6-diphosphate is marked with *.}
\end{center}
\end{figure}

If the formose reaction took place in a phosphate environment with a phosphorylation  before the condensation 
 it explains not only the diversity of the enantiomeric form of the membrane molecules but also
the central step in the glycolysis and gluconeogenesis. In this step 
 dihydroxyacetone phosphate (DHAP) condensates with D-glyceraldehyde-3-phosphate
(D-GLA-3-P) to D-fructose-1,6-diphosphate with the release
of one water molecule (Figure 5).  DHAP is one of
the molecules in the isomerization reaction of  D,L-glyceraldehyde-3-phosphate  and will appear together
with  D-glyceraldehyde-3-phosphate. This aldol condensation 
 is very exergonic at standard concentrations (molar concentrations at pH=7)
 with a Gibbs free energy $\Delta G^{^{\ominus '}}$= -23.98 kJ/mole and the reactants will react spontaneously \cite{Lehninger}.
The reason why we obtain energy  from the reverse
reaction with  hydrolysis of  D-fructose-1,6-diphosphate in the glycolysis   is
 that the concentration  of D-fructose-1,6-diphosphate  is usual quite low ($<$0.1 mM) by
which the reverse reaction is entropy driven and  exergonic. 

\section{In summary}
A break of symmetry and the ordering of a racemic solution into one homochiral domain only requires a strong
chiral discrimination and an  isomerization kinetics. These two conditions can be present in a concentrated
racemic solution.
 The requirement of a high concentration of the chiral reactant(s)
in an aqueous solution  in order to perform and $\textit{maintain}$ homochirality;
the appearance of phosphorylation of almost all carbohydrates in the central machinery of life; the
basic ideas that the biochemistry and  the glycolysis  and  gluconeogenesis contains the trace of the biochemical evolution \cite{Romano,Say},
 all point in the direction of that
homochirality was obtained just after- or at a phosphorylation  of the very first products of the formose reaction
at high concentrations of the reactants in  phosphate-rich compartments.

An alkaline submarine hydrothermal mound is known to be compartmentalized and the
inorganic membranes would offer low water activity traps and sites not only necessary for the  phoshporylations \cite{Russellnew} 
but also for the  domain catalyzed and self-stabilizing 
mechanism by which the homochirality  of simple  organic chiral molecules can be obtained
and maintained over long periods of time.

 Although there have been many experimental
investigations of possible spontaneous ordering of racemic systems into
 homochiral systems there do, however, not exist such experiments for
D,L-glyceraldehyde-3-phosphate(aq).

\begin{acknowledgments}
The author acknowledge useful discussions with Axel Hunding and Jeppe C Dyre.
The centre for viscous liquid dynamics ``Glass and Time'' is sponsored by the Danish National Research Foundation (DNRF) grant No. DNRF61.
\end{acknowledgments}

 \bibliographystyle{spmpsci}      


\end{document}